\documentclass[a4papers,showpacs, showkeys, 12pt]{revtex4}
\usepackage{amsfonts}
\usepackage{graphicx}
\usepackage{amsmath}
\usepackage{amssymb}
\usepackage{latexsym}
\usepackage{psfrag}
\usepackage{anysize}
\usepackage[bookmarks, colorlinks=true, plainpages = false,citecolor 
= blue, urlcolor = black, filecolor = blue]{hyperref}

\begin{document}

\hspace{10 mm}\\
\hspace{10 mm}\\
\hspace{10 mm}\\
\hspace{30 mm}\\

\title{\textbf{Theoretical Investigation of Two-Dimensional Superconductivity in Intercalated Graphene Layers}}

\author{R. A. Jishi}
\email{raljish@exchange.calstatela.edu}
\affiliation{Department of Physics and Astronomy, California State University, Los Angeles, California 90032, USA}

\author{D. M. Guzman}
\email{dguzman4@calstatela.edu}
\affiliation{Department of Physics and Astronomy, California State University, Los Angeles, California 90032, USA}

\author{H. M. Alyahyaei}
\email{halya001@ucr.edu}
\affiliation{Department of Physics and Astronomy, University of California, Riverside, California 92521, USA}



\begin{abstract}
First-principles calculations of the electronic structure and vibrational modes, in a system of graphene bilayers and trilayers intercalated with alkaline earth atoms, are presented. It is found that, in similarity to the case of superconducting graphite intercalation compounds, the Fermi level is crossed by an s-band derived from the intercalant states, as well as graphitic $\pi$-bands. The electron-phonon coupling parameter $\lambda$ is found to be $0.60$ and $0.80$, respectively, in calcium intercalated graphene bilayers and trilayers.  In superconducting CaC$_6$ graphite intercalation compound, the calculated value for $\lambda$ is $0.83$. It is concluded that two-dimensional superconductivity is possible in a system of a few graphene layers intercalated with calcium. 
\end{abstract}


\maketitle

\section{Introduction}
\label{sec:intro}
Superconductivity in graphite intercalation compounds (GICs) has been studied extensively for a long period of time \cite{Hannay, Koike, Kobayashi, Alexander, Iye, Al-Jishi0, Emery1, Weller, Emery2, Walters}. A fascinating aspect of these superconducting compounds is that they are usually formed from components neither one of which is a superconductor. In the case of alkali metal GICs the superconducting transition temperature is of the order of $1$ K. Within the GIC Brillouin zone (BZ) there exists an alkali-derived spherical band centered at the $\Gamma$-point and carbon-derived cylindrical $\pi$-bands centered along the $HK$ axis. It was proposed \cite{Al-Jishi0, Al-Jishi1, Al-Jishi2} that the presence of both bands is necessary for superconductivity, and that a two-band model is capable of explaining various experimental observations relating to the occurrence of superconductivity in alkali-metal GICs and their anisotropic magnetic properties.

It was also found that alkaline earth GICs exhibit superconductivity \cite{Emery1, Weller} with a transition temperature of $6.5$ K (YbC$_6$) and $11.5$ K (CaC$_6$). Careful analysis \cite{Calandra1, Calandra2, Calandra3, Sanna} of the electron-phonon coupling in CaC$_6$ reveals that this superconductor is of the conventional BCS type, and that the presence of both Ca-derived and C-derived bands is necessary to explain the relatively high superconducting transition temperature in this compound. It should be pointed out, however, that some controversy still exists regarding the mechanism for superconductivity in CaC$_6$ GIC, whereas a large Ca isotope effect was measured in CaC$_6$ GIC \cite{Hinks, Gonnelli} indicating a dominant role played by phonons due to Ca atomic vibrations, angle-resolved photoemission spectroscopy (ARPES) measurements on CaC$_6$ GIC \cite{Valla}, on the other hand, indicate that graphitic high frequency phonon modes are most strongly coupled to electrons.

The recent discovery of stable graphene sheets \cite{Novoselov} raises the question of whether two-dimensional superconductivity is possible in intercalated graphene layers. The possibility of intercalating a system consisting of a few graphene layers with intercalants, such as FeCl$_3$ and Br, has been recently demonstrated \cite{Jung1, Jung2}. It is clear that a system consisting of many graphene layers, if intercalated, should exhibit properties similar to those of GICs. It is natural to ask whether a system consisting of a few graphene layers, intercalated with alkali, or alkaline earth, metal atoms, will exhibit superconductivity. This problem was discussed recently by Mazin and Balatsky \cite{Mazin}, who considered a graphene bilayer intercalated with Ca. They argued, on the basis of similarities with the electronic energy bands of the CaC$_6$ GIC, that a graphene bilayer intercalated with Ca is probably a superconductor.

In this paper we examine the problem of superconductivity in intercalated graphene layers. In particular, we calculate the energy bands in a graphene bilayer intercalated with the alkaline earth atoms Ca, Sr, and Ba. In all three cases we find that a band, derived from the alkaline earth atoms, crosses the Fermi energy, along with carbon-derived $\pi$-bands. We then calculate the electron-phonon coupling strength $\lambda$ in these systems. The values obtained are reasonably large; in the case of Ca-intercalated bilayer, we find $\lambda=0.60$, a value to be compared with $0.83$ calculated in CaC$_6$ GIC \cite{Calandra0}. Next, the case of a graphene trilayer intercalated with Ca is examined; here we find a value of $0.80$ for the electron-phonon coupling strength $\lambda$. These calculations indicate that superconductivity is indeed possible in intercalated graphene layers.

In Section \ref{sec:meths} we discuss the computational methods employed in this work. The results of the energy bands, vibrational modes, and electron-phonon coupling calculations are presented and discussed in Section \ref{sec:R&D}, and conclusions are given in Section \ref{sec:con}.

\section{Methods}
\label{sec:meths}
The total energy and electronic band structure calculations are carried out using the all-electron, full-potential, linear augmented plane wave (FP-LAPW) method as implemented in the WIEN2K code \cite{WIEN2K}. The exchange-correlation potential was calculated using the generalized gradient approximation (GGA) as proposed by Pedrew, Burke, and Ernzerhof (PBE) \cite{PBE}. The radii of the muffin-tin spheres for the carbon and the alkaline earth atoms were taken as $1.3a_0$ and $2.5a_0$, respectively, where $a_0$ is the Bohr radius. We set the parameter $R_{MT}K_{max}=7$, where $R_{MT}$ is the smallest muffin-tin radius and $K_{max}$ is a cutoff wave vector. The valence electrons wave functions inside the muffin-tin spheres are expanded in terms of spherical harmonics up to $l_{max}=10$, and in terms of plane waves with a wave vector cutoff $K_{max}$ in the interstitial region. The charge density is Fourier expanded up to a maximum wave vector $G_{max}=13a_0^{-1}$. Convergence of the self-consistent field calculations is attained with a total energy convergence tolerance of $0.01$ mRy.

The calculations of the frequencies of the vibrational modes and the electron-phonon coupling parameter are performed using density functional theory (DFT) and ultrasoft pseudopotentials \cite{QE1, QE2}. The electron-phonon coupling strengths are calculated using density functional perturbation theory within the linear response approximation. Ultrasoft pseudopotentials \cite{Vanderbilt} are used for both, the alkaline earth and the carbon atoms. The valence electrons wave function and charge density are expanded in plane waves using $30$ Ry and $300$ Ry cutoffs, respectively. In the calculation of the vibrational frequencies the electronic integration is carried out using a uniform mesh of $N_k=8\times 8$ k-points in the two-dimensional BZ. On the other hand, the electronic density of states, used in computing the electron-phonon coupling parameter $\lambda$, is calculated using a finer uniform mesh of $24\times 24$ k-points. The value of $\lambda$ is obtained by averaging over a uniform mesh of $N_q=6\times 6$ phonon momentum q-points.

\section{Results and Discussion}
\label{sec:R&D}
The first issue to decide is this: given a graphene bilayer, will it be energetically favorable for the alkaline earth atoms to be intercalated between the two carbon sheets comprising the bilayer? Or will the energy be lower if the alkaline earth atoms sit on top of the bilayer? If the atoms sit on top of the bilayer, the resulting structure will be denoted by C$_{6}$-C$_{6}$-A, where A represents an alkaline earth atom. On the other hand, the intercalated structure is denoted by C$_{6}$-A-C$_{6}$. In order to decide which structure is more stable, we carried out total energy calculations for both structures. In the case of C$_6$-C$_6$-A, the alkaline earth atom is placed over the center of a hexagon in the upper sheet, and its distance to the sheet is varied in order to find the lowest energy. On other hand, in the case of C$_{6}$-A-C$_{6}$, the distance between the carbon sheets is fixed at the measured separation in the corresponding graphite intercalation compound \cite{Emery2, Guerard}. In all the cases considered, corresponding to A=Ca, Sr, and Ba, we found that the energy is lower for the intercalated structure. For the case A=Ca, the intercalated structure C$_{6}$-Ca-C$_{6}$ is lower in energy than the C$_{6}$-C$_{6}$-Ca structure by $0.47\ eV/$Ca atom. In the other two cases, A=Sr and Ba, the energy of the C$_{6}$-A-C$_{6}$ structure is lower than the C$_{6}$-C$_{6}$-A structure by $0.43\  eV/$Sr atom and $0.59\ eV/$Ba atom. We conclude that it is energetically favorable for the alkaline earth atom to be intercalated in between the graphene layers. 

\begin{figure}[htbp]
    \includegraphics[scale=0.4]{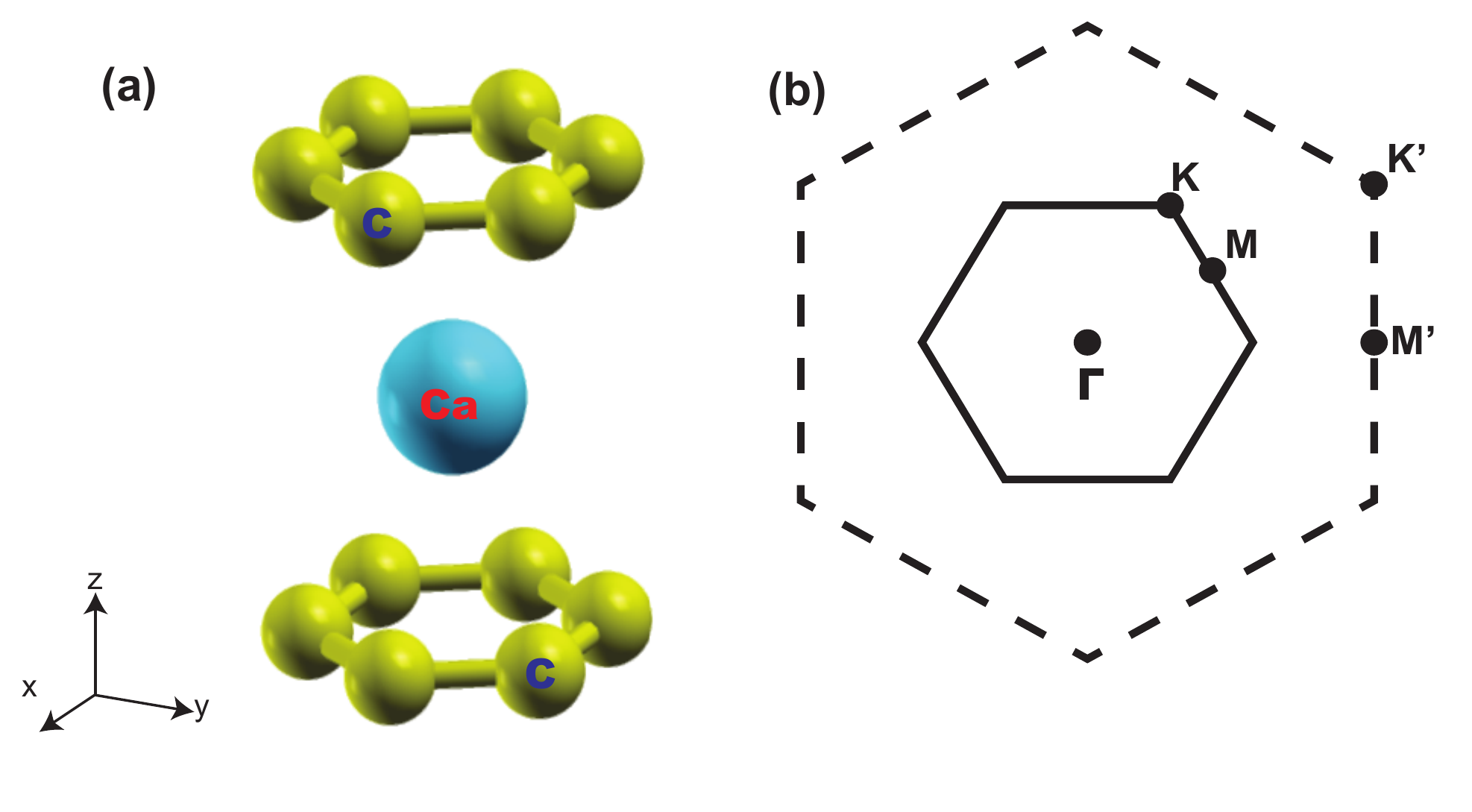}
\caption{(Color online) (a) The unit cell of the structure C$_6$-A-C$_6$, where A is an alkaline earth atom, is shown. (b) The 2-D Brillouin zone of a graphene layer (or bilayer) is shown as a dashed hexagon, while that for the intercalated graphene bilayer is shown as a solid hexagon. Note that the M and K points in the graphene BZ (here denoted by M' and K') are folded onto points M and $\Gamma$, respectively, of the intercalated bilayer BZ.}
\label{fig:unitcell}
\end{figure}

In the following we focus attention only on the C$_6$-A-C$_6$ structure. The separation between the two carbon sheets of the intercalated bilayer is taken as $4.52$ \AA, $4.94$ \AA, and $5.25$ \AA\  for A=Ca, Sr, and Ba, respectively. The unit cell is a $(\sqrt{3}a \times \sqrt{3}a)$ $R$ $30^{\circ}$ supercell of the graphene sheet unit cell; i.e., the in-plane lattice constant of C$_{6}$-A-C$_{6}$ is $\sqrt{3}$ times the lattice constant in the graphene sheet, and the in-plane lattice vectors of C$_{6}$-A-C$_{6}$ are rotated by $30^{\circ}$ relative to those in the graphene sheet. The intercalated bilayer unit cell, along with the first Brillouin zones of the graphene sheet and the supercell structure are shown in Figure \ref{fig:unitcell}. Since the DFT calculations are carried out on a $3$-dimensional crystal, the $c$-lattice constant is taken to be sufficiently large ($c=14.6$ \AA, $15.0$ \AA, and $15.3$ \AA\ for A= Ca, Sr, and Ba, respectively) so that adjacent intercalated bilayers have negligible interactions.

\begin{figure}[htbp]
    \includegraphics[scale=0.5]{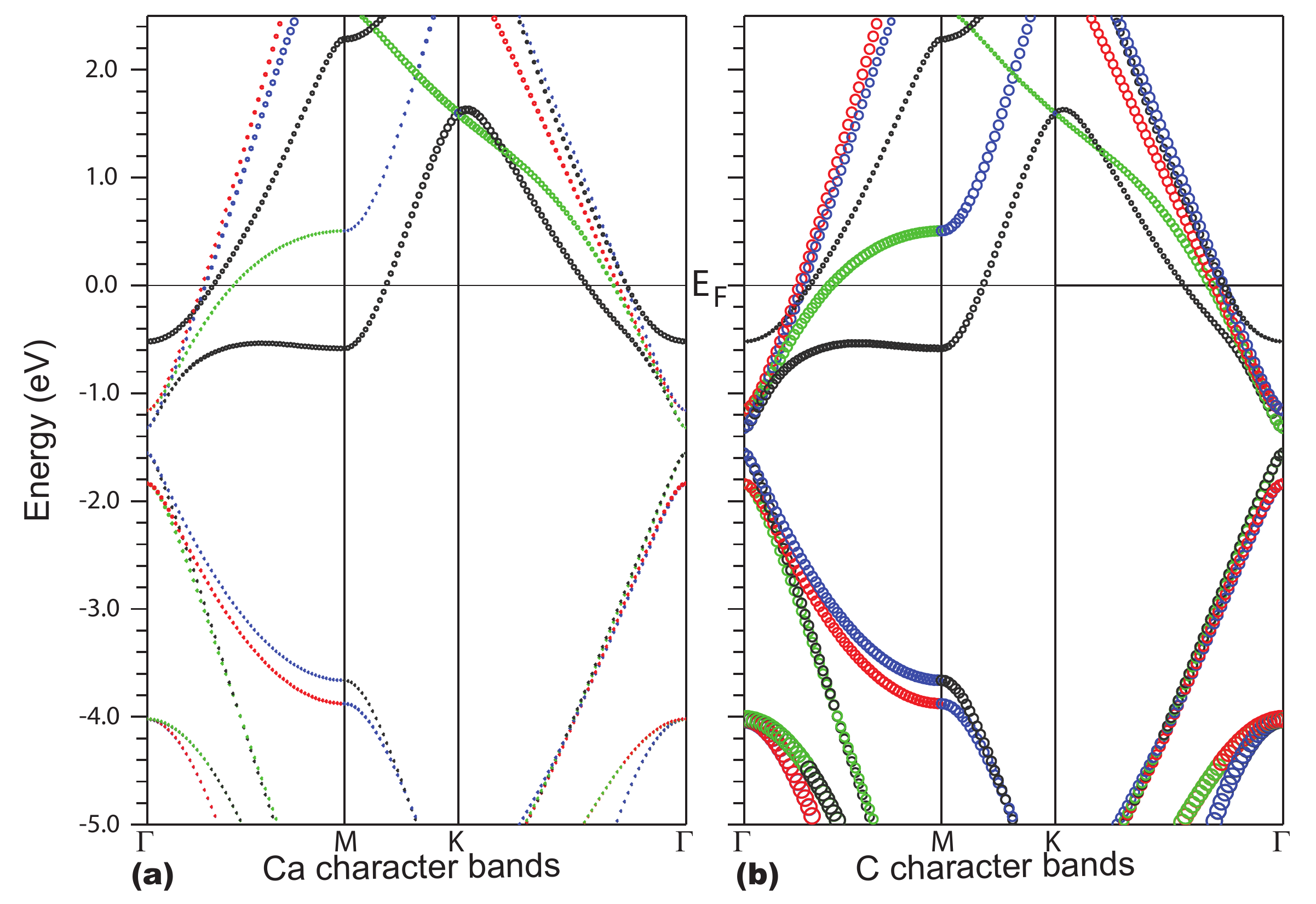}
\caption{(Color online) Plot of the energy bands in Ca-intercalated graphene bilayer showing the character of the contributing states. The width of the band at every k-point is proportional to the contribution of (a) Ca-derived states, and (b) C-derived states.}
\label{fig:bands}
\end{figure}

\begin{figure}[htbp]
    \includegraphics[scale=0.5]{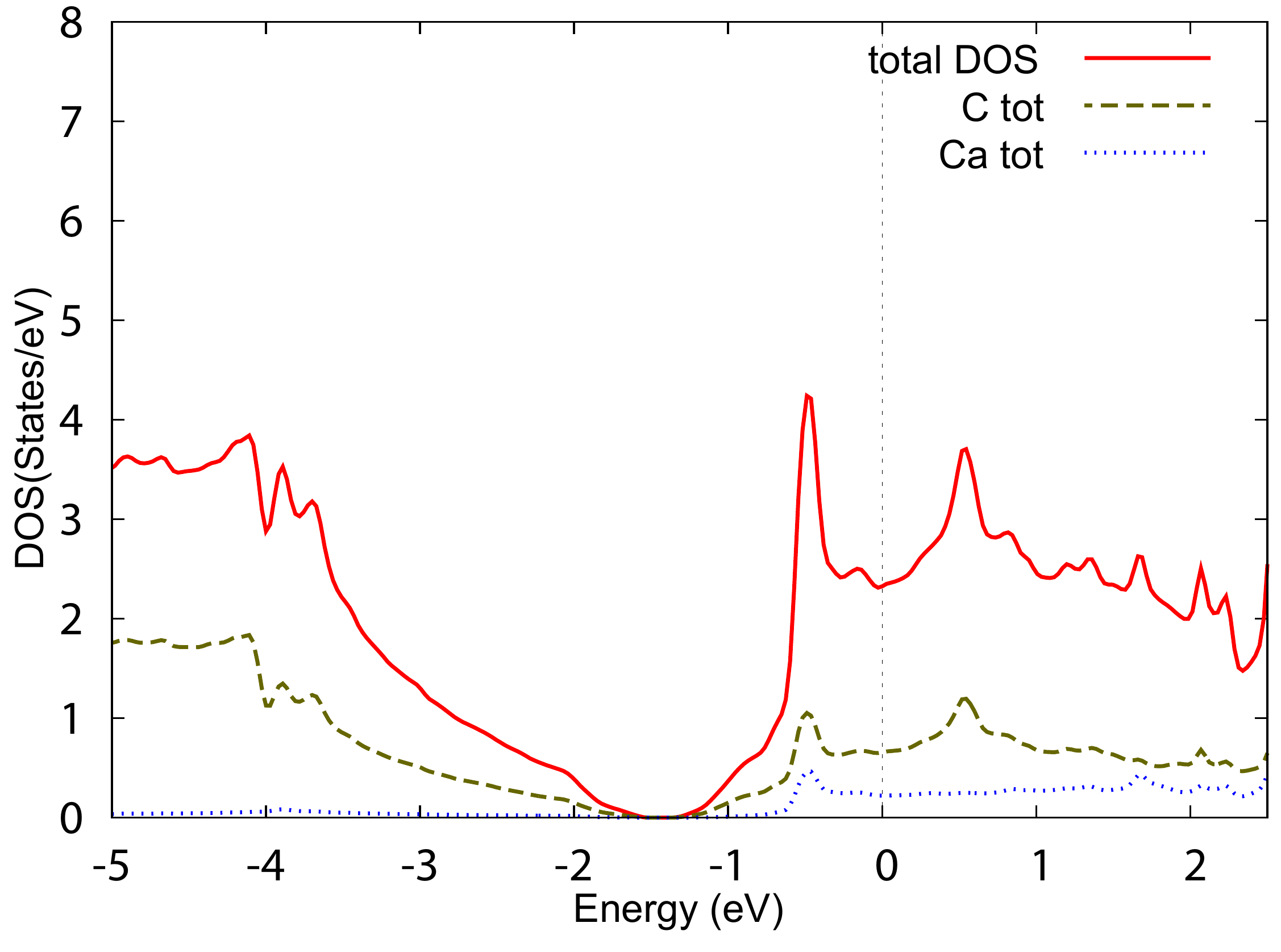}
\caption{(Color online) Total and atomic densities of states in Ca-intercalated graphene bilayer. The total density of states is the sum of the atomic densities of states and the density of states in the interstitial region. The Fermi energy is at E$=0$.}
\label{fig:DOS}
\end{figure}

In Figure \ref{fig:bands} the calculated energy bands in C$_{6}$-Ca-C$_{6}$ are plotted in the fatband representation in two different panels, along high symmetry directions in the first BZ. The energy band dispersion curves in both panels are identical, but the width (or the ``fatness'') of the band at every k-point in Figure \ref{fig:bands}a is proportional to the contribution of the Ca atoms to the band eigenstates, while in Figure \ref{fig:bands}b, the width is proportional to the contribution of the C-derived states. We note that an s-band originating from the Ca states, as well as $\pi$-bands originating from C states, cross the Fermi level. This is indicative of the partial charge transfer from the intercalant atoms to the carbon sheets. Indeed, using the Bader's atoms in molecule (AIM) method\cite{Bader}, we calculated a charge transfer to the carbon sheets of $1.2$ electrons/Ca atom, $1.06$ electrons/Sr atom, and $0.91$ electrons/Ba atom. We also note that in the $(\sqrt{3}\times\sqrt{3})$ $R$ $30^{\circ}$ supercell structure, the K-point of the BZ of the graphene sheet is folded onto the $\Gamma$ point of the BZ of the supercell, as Figure \ref{fig:unitcell} indicates. This explains the observation that the $\pi$-bands that are centered at the K-point in graphene are now centered at the $\Gamma$-point in the band structure shown in Figure \ref{fig:bands}. The total and atomic densities of states are plotted in Figure \ref{fig:DOS}. We note that the total density of states (DOS) consists of the sum of the atomic DOS and the DOS in the interstitial region. The plot in Figure \ref{fig:DOS} reveals that there is a substantial contribution to the total DOS coming from the interstitial region.

It has been argued previously that for superconductivity in GICs to exist, it is necessary that the charge transfer from the intercalant to carbon sheets be incomplete \cite{Al-Jishi1, Al-Jishi2}, so that a band derived from the intercalant states must coexist with the $\pi$-bands derived from the carbon states. Superconductivity was then explained as resulting from scattering of electrons, through phonon emission or absorption, between the intercalant and the $\pi$-bands. More recently, DFT calculations showed that in the CaC$_6$ GIC, the $\pi$-bands alone can not account for the observed superconducting transition temperature, and that the presence of the Ca-derived band is necessary \cite{Calandra1, Calandra2, Sanna, Calandra3}. Hence, it is reasonable to expect that superconductivity is possible in the intercalated bilayers under discussion in this work. The superconducting transition temperature will be determined mainly by the strength of the electron-phonon coupling in these systems.

Similar to the case of CaC$_{6}$ GICs \cite{Calandra0}, we calculate the phonon spectrum of the bilayers using DFT as implemented in the PWSCF code \cite{QE1, QE2}. The technical details of the calculation are given in Section \ref{sec:meths}.

For the graphene bilayer intercalated with alkaline earth atoms, the crystal point group is $D_{6h}=D_6 \times i$, where $i$ is the inversion symmetry. The group's character table is given in Table \ref{tab:character}. The C$_6$ axis is the $z$-axis which passes through the alkaline earth atom and is perpendicular to the carbon sheets. The $C'_2$ axes in the $x-y$ plane are parallel to the lines bisecting the hexagonal sides, and the $C''_2$ axes are parallel to the hexagonal diagonals. The $3N=39$-dimensional representation, $\Gamma$, is given at the bottom of Table \ref{tab:character}; it is obtained by considering the transformation properties of the $39$ cartesian unit vectors placed on the thirteen atoms of the unit cell. From Table \ref{tab:character}, the $39$ phonon modes at the BZ center are decomposed into irreducible representations according to
\begin{eqnarray*}
\Gamma_{phonon} &=& 2A_{1g} + A_{2g} + 2B_{1g} + B_{2g} + 3E_{1g} + 3E_{2g}\\
                &+& A_{1u} + B_{1u} + 3A_{2u} + 2B_{2u} + 4E_{1u} + 3E_{2u}
\end{eqnarray*}

 \begin{table}[ht]
       \caption{\label{tab:character}Character table of $D_{6h}=D_6 \times i$. The character of the 39-dimensional representation, $\Gamma$, is given at the bottom.}
        \begin{ruledtabular}
         \begin{tabular}{c|c|c c c c c c| c c c c c c}
     Axes   & Modes  &$E$&$C_2$  &$2C_3$ &$2C_6$ &$3C'_2$&$3C''_2$&$i$    &$iC_2$ &$2iC_3$&$2iC_6$&$3iC'_2$&$3iC''_2$   \\ \hline        
            &$A_{1g}$&$1$&$\ \ 1$&$\ \ 1$&$\ \ 1$&$\ \ 1$&$\ \ 1$ &$\ \ 1$&$\ \ 1$&$\ \ 1$&$\ \ 1$&$\ \ 1$ &$\ \ 1$ \\
            &$A_{1u}$&$1$&$\ \ 1$&$\ \ 1$&$\ \ 1$&$\ \ 1$&$\ \ 1$ &$-1$   &$-1$   &$-1$   &$-1$   &$-1$    &$-1$    \\ 
            &$A_{2g}$&$1$&$\ \ 1$&$\ \ 1$&$\ \ 1$&$-1$   &$-1$    &$\ \ 1$&$\ \ 1$&$\ \ 1$&$\ \ 1$&$-1$    &$-1$    \\  
     $z$    &$A_{2u}$&$1$&$\ \ 1$&$\ \ 1$&$\ \ 1$&$-1$   &$-1$    &$-1$   &$-1$   &$-1$   &$-1$   &$\ \ 1$ &$\ \ 1$ \\
            &$B_{1g}$&$1$&$-1$   &$\ \ 1$&$-1$   &$\ \ 1$&$-1$    &$\ \ 1$&$-1$   &$\ \ 1$&$-1$   &$\ \ 1$ &$-1$    \\
            &$B_{1u}$&$1$&$-1$   &$\ \ 1$&$-1$   &$\ \ 1$&$-1$    &$-1$   &$\ \ 1$&$-1$   &$\ \ 1$&$-1$    &$\ \ 1$ \\
            &$B_{2g}$&$1$&$-1$   &$\ \ 1$&$-1$   &$-1$   &$\ \ 1$ &$\ \ 1$&$-1$   &$\ \ 1$&$-1$   &$-1$    &$\ \ 1$ \\
            &$B_{2u}$&$1$&$-1$   &$\ \ 1$&$-1$   &$-1$   &$\ \ 1$ &$-1$   &$\ \ 1$&$-1$   &$\ \ 1$&$\ \ 1$ &$-1$    \\
            &$E_{1g}$&$2$&$-2$   &$-1$   &$\ \ 1$&$\ \ 0$&$\ \ 0$ &$\ \ 2$&$-2$   &$-1$   &$\ \ 1$&$\ \ 0$ &$\ \ 0$ \\
 $(x,y)$    &$E_{1u}$&$2$&$-2$   &$-1$   &$\ \ 1$&$\ \ 0$&$\ \ 0$ &$-2$   &$\ \ 2$&$\ \ 1$&$-1$   &$\ \ 0$ &$\ \ 0$ \\
            &$E_{2g}$&$2$&$\ \ 2$&$-1$   &$-1$   &$\ \ 0$&$\ \ 0$ &$\ \ 2$&$\ \ 2$&$-1$   &$-1$   &$\ \ 0$ &$\ \ 0$ \\
            &$E_{2u}$&$2$&$\ \ 2$&$-1$   &$-1$   &$\ \ 0$&$\ \ 0$ &$-2$   &$-2$   &$\ \ 1$&$\ \ 1$&$\ \ 0$ &$\ \ 0$ \\    \hline    
            &$\Gamma$&$39$&$-1$  &$\ \ 0$    &$\ \ 2$    &$-1$   &$-1$    &$-3$   &$\ \ 1$    &$\ \ 0$    &$-2$   &$\ \ 5$     &$\ \ 1$\\ 
\end{tabular}

        \end{ruledtabular}
\end{table}

The $A_{1g}$, $E_{1g}$, and $E_{2g}$ modes are Raman-active while the $A_{2u}$ and $E_{1u}$ modes are infrared-active. From the representations of $x$, $y$, and $z$, in Table \ref{tab:character}, the three translational modes are found to transform as $A_{2u}$ and $E_{1u}$. The Raman and infrared-active modes, with nonzero frequencies, for the three intercalated bilayers considered in this work are given in Table \ref{tab:bimodes}. The overall trend displayed in Table \ref{tab:bimodes} is in accordance with what is expected on general grounds. The high frequency modes are lower than the corresponding modes in graphite as a result of the considerable charge transfer which softens the force constants in the carbon sheets. The frequencies of the higher modes increase slightly as the intercalant is changed from Ca to Sr to Ba, a consequence of the concomitant decrease in the amount of charge transfer. The validity of the first-principles calculations of the phonon spectra in the intercalated bilayer may be checked by repeating the same calculation for pristine graphite where the $\Gamma$ point phonon frequencies are known experimentally. In graphite, the calculated phonon frequencies of the Raman-active modes are $\omega(E_{2g})=75$ cm$^{-1}$ and $1557$ cm$^{-1}$, the IR-active modes $\omega(A_{2u})=892$ cm$^{-1}$, $\omega(E_{1u})=1567$ cm$^{-1}$, and the silent modes $\omega(B_{1g})=133$ cm$^{-1}$ and $889$ cm$^{-1}$. Experimentally, the measured frequencies \cite{Tunista, Brillson, Nemanich, Nicklow} of the Raman-active modes are $42$ and $1582$ cm$^{-1}$, the IR-active modes $868$ and $1588$ cm$^{-1}$, and the low frequency silent mode $128$ cm$^{-1}$. It is seen that, except for the low frequency mode at 42 cm$^{-1}$, a good agreement exists between the first-principles calculations and experiment, the error being of the order of a few percent. 

\begin{table}[ht]
       \caption{\label{tab:bimodes}Raman- and Infrared-active modes in the intercalated graphene bilayer C$_{6}$-A-C$_{6}$ for A=Ca, Sr, and Ba.}
        \begin{ruledtabular}
         \begin{tabular}{l|c|c c c c}
        \multicolumn{3}{c}{\ } & \multicolumn{2}{c}{$\omega$(irrep)/cm$^{-1}$}&\multicolumn{1}{c}{\ }\\ \hline
         Raman-active modes  & Ca & 69($E_{1g}$)   & 128($A_{1g}$)  & 497($E_{1g}$) & 498($E_{2g}$) \\
                            &    & 1132($E_{2g}$) & 1135($E_{1g}$) & 1343($A_{1g}$)& 1410($E_{2g}$) \\ 
                            & Sr & 45($E_{1g}$)   & 133($A_{1g}$)  & 506($E_{1g}$) & 511($E_{2g}$) \\
                            &    & 1152($E_{2g}$) & 1152($E_{1g}$) & 1355($A_{1g}$)& 1413($E_{2g}$) \\
                            & Ba & 38($E_{1g}$)   & 115($A_{1g}$)  & 506($E_{1g}$) & 509($E_{2g}$)  \\
                            &    & 1166($E_{1g}$) & 1167($E_{2g}$) & 1369($A_{1g}$)& 1454($E_{2g}$)  \\\hline
     Infrared-active modes  & Ca & 115($E_{1u}$)  & 293($A_{2u}$)  & 481($E_{1u}$) & 1130($E_{1u}$) \\
                            &    & 1349($A_{2u}$) &                &               &                \\
                            & Sr & 72($E_{1u}$)   & 207($A_{2u}$)  & 498($E_{1u}$) & 1151($E_{1u}$) \\
                            &    & 1358($A_{2u}$) &                &               &                \\
                            & Ba & 67($E_{1u}$)   & 153($A_{2u}$)  & 503($E_{1u}$) & 1165($E_{1u}$) \\
                            &    & 1370($A_{2u}$) &                &               &                \\ 
  \end{tabular}
        \end{ruledtabular}

\end{table}

We calculated the electron-phonon coupling parameter $\lambda$ in C$_6$-A-C$_6$ and found it to be $0.60$, $0.42$, and $0.35$, for A=Ca, Sr, and Ba respectively. The transition superconducting temperature T$_c$ of the the studied compounds was estimated using the Allen-Dynes \cite{Allen} formula with values for the phonon logarithmic average frequencies $\omega_{log}=29.4$ meV, $30.1$ meV and $28.4$ meV for C$_6$-Ca-C$_6$, C$_6$-Sr-C$_6$, and C$_6$-Ba-C$_6$ respectively. Using a screened Coulomb pseudopotential $\mu^*=0.14$ we obtained T$_c=4.7$ K for Ca-intercalated graphene bilayer, $0.95$ K for Sr-intercalated bilayer, and $0.1$ K for Ba-intercalated bilayer. This is to be compared with the value $\lambda=0.83$ calculated previously in CaC$_6$ GIC \cite{Calandra0} for which T$_c$ is $11.5$ K.   

\begin{figure}[htbp]
    \includegraphics[scale=0.6]{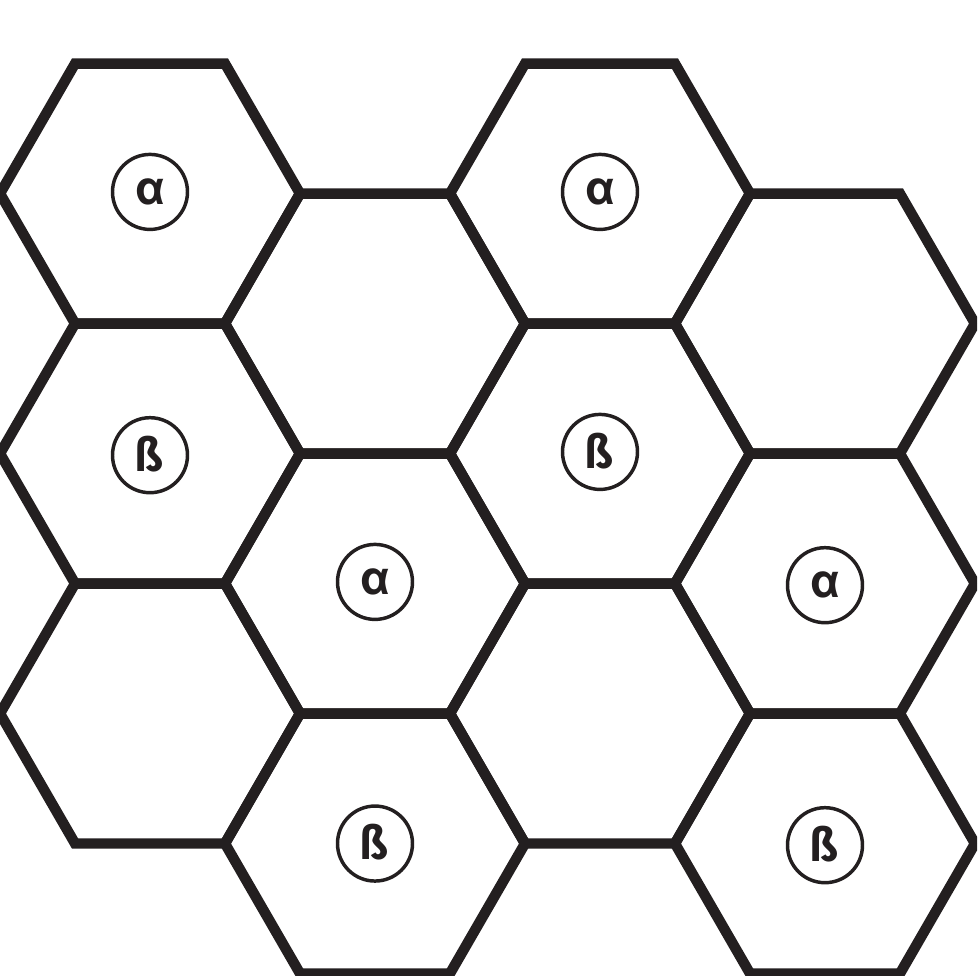}
\caption{(Color online) Projection, on the plane of the first carbon layer, of the intercalated graphene trilayer. In the first intercalant layer, Ca atoms occupy positions denoted by $\alpha$, whereas Ca atoms occupy positions $\beta$ in the second intercalant layer.}
\label{fig:trilayer}
\end{figure}

In order to study the effect of increasing number of graphene layers on the superconducting properties, we consider a trilayer of graphene intercalated with calcium: Ca$_2$C$_{18}$. The unit cell is again given by $(\sqrt{3}a \times \sqrt{3}a)\ R\ 30^{\circ}$ supercell of the graphene sheet unit cell. In conformity with the corresponding situation in stage-1 GICs, the carbon layer stacking is taken to be AAA. Between the first and second carbon layers, Ca atoms sit above hexagon centers, occupying positions labeled $\alpha$. Between the second and third carbon layers, Ca atoms sit above hexagon centers, occupying positions labeled $\beta$. The two intercalant layers are not directly above each other; the second intercalant layer is shifted, relative to the first intercalant layer, by a lattice vector of the graphene sheet, as illustrated in Figure \ref{fig:trilayer}. Thus, the stacking in the intercalated graphene trilayer is A$\alpha$A$\beta$A. The crystal point group is reduced, from $D_{6h}$ in the case of intercalated bilayer, to $D_{3d}$ in the case of the intercalated trilayer. The phonon modes, at the BZ center, are decomposed into the following irreducible representations
\begin{eqnarray*}
\Gamma_{phonon} &=& 5A_{1g} + 5A_{2g} + 4A_{1u} + 6A_{2u} + 10E_{g} + 10E_{u}.
\end{eqnarray*}

\begin{table}[htbp]
       \caption{\label{tab:trimodes}Raman- and Infrared-active modes, with nonzero frequencies, in Ca-intercalated graphene trilayer.}
        \begin{ruledtabular}

         \begin{tabular}{l|c c c c c}
        \multicolumn{3}{c}{\ } & \multicolumn{2}{c}{$\omega$(irrep)/cm$^{-1}$}&\multicolumn{1}{c}{\ }\\ \hline
         Raman-active modes    & 40($E_{g}$)    & 83($A_{1g}$)    & 102($E_{g}$)   & 249($A_{1g}$)  & 411($E_{g}$) \\
                               & 489($E_{g}$)   & 492($E_{g}$)   & 1112($E_{g}$)  & 1131($E_{g}$)   & 1133($E_{g}$) \\ 
                               & 1326($A_{1g}$) & 1344($A_{1g}$) & 1362($E_{g}$)  & 1382($A_{1g}$) & 1440($E_{g}$) \\ \hline
      Infrared-active modes    & 65($E_{u}$)    & 106($E_{u}$)   & 159($A_{2u}$)  & 285($A_{2u}$)  & 425($E_{u}$) \\
                               & 489($E_{u}$)   & 492($E_{u}$)   & 1111($E_{u}$)  & 1132($E_{u}$)  & 1133($E_{u}$) \\
                               & 1326($A_{2u}$) & 1344($A_{2u}$) & 1362($E_{u}$)  & 1401($A_{2u}$)\\ 
         \end{tabular}
        \end{ruledtabular}

\end{table} 
The $A_{1g}$ and $E_{g}$ modes are Raman-active, whereas $A_{2u}$ and $E_{u}$ modes are IR-active. The three zero-frequency translational modes transform according to $A_{2u}$ and $E_{u}$ irreducible representations. In Table \ref{tab:trimodes} we report the Raman- and IR-active modes with non-vanishing frequencies. The electron-phonon coupling parameter in calcium intercalated graphene trilayer is calculated to be $\lambda=0.80$. The calculated phonon logarithmic average frequency is $25.2$ meV, and the calculated superconducting transition temperature is $10.1$ K. 


\section{Conclusions}
\label{sec:con}
We have carried out energy band calculations, within density functional theory, of a graphene bilayer intercalated with the alkaline earth atoms A=Ca, Sr, or Ba, and a graphene trilayer intercalated with Ca. In all cases, we found that the Fermi level is crossed by an s-band derived from the intercalant states as well as graphitic $\pi$-bands, in similarity to the situation in superconducting graphite intercalation compounds. The electron-phonon coupling parameter $\lambda$ is largest for the case A=Ca, where $\lambda=0.60$ for the intercalated bilayer, and $\lambda=0.80$ for the intercalated trilayer; this is to be compared with $\lambda=0.83$ for the case of CaC$_6$ GIC. On the basis of these calculations, we expect that graphene bilayers and trilayers, intercalated with calcium, will exhibit two-dimensional superconductivity. Even though there are no experimental reports yet of the successful intercalation of calcium into a system of a few graphene layers, it is hoped that this work will stimulate further experimental studies in that direction


\begin{acknowledgements}
One of the authors (R. A. J) gratefully acknowledges support by NSF under grant No. HRD-0932421. D. M. G. would also like to thank the CSULA-MORE programs and CSU-LSAMP fellowship for their support under grant No. HRD-1026102-518741.   
\end{acknowledgements}



\end{document}